\documentclass[aps, prd, twocolumn, superscriptaddress, nofootinbib]{revtex4}

\usepackage{bm,graphicx}
\usepackage{dcolumn}
\usepackage{amssymb}

\begin{document}

\newcommand{\fs}{f_{10}}
\newcommand{\fd}{f_{10}}
\newcommand{\As}{A_{\mathrm{s}}}
\newcommand{\At}{A_{\mathrm{t}}}
\newcommand{\ns}{n_{\mathrm{s}}}
\newcommand{\nt}{n_{\mathrm{t}}}
\newcommand{\Obhh}{\Omega_{\mathrm{b}}h^{2}}
\newcommand{\Omhh}{\Omega_{\mathrm{m}}h^{2}}
\newcommand{\Ol}{\Omega_{\Lambda}}
\newcommand{\VEV}{\phi_{0}}
\newcommand{\Sr}{^{\mathrm{S}}}
\newcommand{\Tr}{^{\mathrm{T}}}
\newcommand{\Vr}{^{\mathrm{V}}}
\newcommand{\conj}{^*}
\newcommand{\vect}[1]{\mathbf{#1}}
\newcommand{\half}{\frac{1}{2}}
\newcommand{\FT}[1]{\tilde{#1}}
\newcommand{\tauOS}{\tau_{\xi=0}}
\newcommand{\CMB}{\textsc{CMB }}
\newcommand{\CMBEASY}{\textsc{CMBeasy }}
\newcommand{\TT}{\textsc{tt }}
\newcommand{\TE}{\textsc{te }}
\newcommand{\EE}{\textsc{ee }}
\newcommand{\BB}{\textsc{bb }}
\newcommand{\WMAP}{\textsc{WMAP }}
\newcommand{\UETC}{\textsc{uetc }}
\newcommand{\UETCs}{\textsc{uetc}s }
\newcommand{\ETC}{\textsc{etc }}
\newcommand{\ETCs}{\textsc{etc}s }
\newcommand{\HK}{\textsc{HKP }}
\newcommand{\BBN}{\textsc{BBN }}
\newcommand{\mcmc}{\textsc{MCMC }}

\newcommand{\rmin}{r_{\mathrm{min}}}
\renewcommand{\d}{{\partial}}
\newcommand{\be}{\begin{equation}}
\newcommand{\ee}{\end{equation}}
\newcommand{\bea}{\begin{eqnarray}}
\newcommand{\eea}{\end{eqnarray}}
\newcommand{\longprd}{{Bevis:2006mj}}
\newcommand{\cmbdata}{{Hinshaw:2006ia,Kuo:2002ua,Jones:2005yb,Readhead:2004gy,Dickinson:2004yr,Page:2006hz}}

\newcommand{\simleq}{\;
  \raisebox{-0.4ex}{\tiny$\stackrel{{\textstyle<}}{\sim}$}\;}   

\title{Cosmic microwave anisotropies from BPS semilocal
strings} 

\newcommand{\addressSussex}{Department of Physics \&
Astronomy, University of Sussex, Brighton, BN1 9QH, United Kingdom}

\author{Jon Urrestilla}
\email{j.urrestilla@sussex.ac.uk}
\affiliation{\addressSussex}

\author{Neil Bevis} 
\email{n.bevis@imperial.ac.uk}
\affiliation{Theoretical Physics, Blackett Laboratory, Imperial
  College, London, SW7 2BZ, United Kingdom} 
\affiliation{\addressSussex}

\author{Mark Hindmarsh} 
\email{m.b.hindmarsh@sussex.ac.uk}
\affiliation{\addressSussex}

\author{Martin Kunz}
\email{martin.kunz@physics.unige.ch}
\affiliation{D\'epartement de Physique Th\'eorique, Universit\'e de
Gen\`eve, 1211 Gen\`eve 4, Switzerland} 
\affiliation{\addressSussex}

\author{Andrew R.~Liddle} 
\email{a.liddle@sussex.ac.uk}
\affiliation{\addressSussex}

\date{11/07/2008}

\begin{abstract}
We present the first ever calculation of cosmic microwave background
(\textsc{CMB}) anisotropy power spectra from semilocal cosmic strings, obtained via
 simulations of a classical field theory.
Semilocal strings are a type of non-topological defect arising in some
models of inflation motivated by fundamental physics, and are thought
to relax the constraints on the symmetry breaking scale
 as compared to models with
(topological) cosmic strings.  We derive constraints on the model
parameters, including the string tension parameter $\mu$, from fits to
cosmological data, and find that in this regard BPS semilocal strings
resemble global textures more than topological strings.  The observed
microwave anisotropy at $\ell = 10$ is reproduced if $G\mu = 5.3
\times 10^{-6}$ ($G$ is Newton's constant). However as with other
defects the spectral shape does not match observations, and in models
with inflationary perturbations plus semilocal strings the $95\%$
confidence level upper bound is $G\mu<2.0\times 10^{-6}$ when \CMB
data, Hubble Key Project and Big Bang Nucleosynthesis data are used
(c.f.\ $G\mu<0.9\times 10^{-6}$ for cosmic strings). We additionally
carry out a Bayesian model comparison of several models with and
without defects, showing models with defects are neither conclusively
favoured nor disfavoured at present.
\end{abstract}

\maketitle


\section{Introduction}

Recent observational data are tightly constraining cosmological
models, in many cases rendering them non-viable. One of the
constraints comes from the fact that many potentially successful
high-energy physics motivated models of inflation predict a cosmic
string network \cite{Vilenkin:1994book,Hindmarsh:1994re} in the
post-inflationary era: this occurs in supersymmetric hybrid inflation
\cite{Lyth:1998xn}, in grand unified theories \cite{Jeannerot:2003qv}
and in some string theory inflation models
\cite{Majumdar:2002hy,Sarangi:2002yt,Copeland:2003bj,Dvali:2003zj}. 
These strings and
other type of defects contribute to the cosmic microwave background
(\textsc{CMB}) anisotropies in addition to the primordial inflationary
fluctuations, allowing \CMB experiments to provide upper limits on the
contribution from defects \cite{Bevis:2004wk,Wyman:2005tu,
Battye:2006pk, Fraisse:2006xc, Bevis:2007gh}. In the near future,
observations will either reveal the presence of these defects or will
significantly tighten the constraints upon the models from which they
are predicted (see for example
Refs.~\cite{Seljak:2006hi,Bevis:2007qz,Battye:2007si,Fraisse:2007nu,Pogosian:2007gi}).

However, a simple and elegant method for evading the constraints
coming from cosmic strings in these models was proposed in
Refs.~\cite{Urrestilla:2004eh,Dasgupta:2004dw}. A duplication of the
matter fields responsible for the string formation, which arises quite
naturally in several models \cite{Dasgupta:2004dw,Achucarro:2005vz,Dasgupta:2007ds},
changes the nature of the strings from topological and
therefore long-lived strings, to non-topological semilocal strings
\cite{Vachaspati:1991dz}.  All the desired inflationary properties of
the model persist, and semilocal strings are believed to be
cosmologically less harmful than their topological counterparts, in
that their energy density may be lower for a given 
symmetry breaking scale.
In this paper we confirm these expectations by performing
the first \CMB calculations for semilocal strings and determine the
constraints upon semilocal strings imposed by current data.

Calculating the additional contribution that cosmic strings would make
to the \CMB temperature and polarization power spectra is
not an easy task. They are highly non-linear entities, and the range
of scales involved in the problem is enormous. 
However, as the width of the string is much smaller
than the string length, one possibility is to neglect the finite width
and perform Nambu--Goto type simulations
\cite{Allen:1996wi,Allen:1997ag,Contaldi:1998mx,Landriau:2003xf}.
Unfortunately, these simulations ignore the string decay into
gravitational waves or particles and the corresponding back-reaction
onto the network. Their representation of small-scale loop production
in the network is also under debate
\cite{Vincent:1996rb,Vanchurin:2005pa,Vanchurin:2005yb,
Ringeval:2005kr,Martins:2005es}. Yet another level of simplification
often used in \CMB calculations for local strings is to simulate
merely a stochastic ensemble of unconnected string segments
\cite{Albrecht:1997nt,Pogosian:1999np,Wyman:2005tu,
Seljak:2006hi,Battye:2006pk,Battye:2007si}.  Lacking any dynamical
equations, in this case the sub-horizon decay of strings must be taken
into account by the random removal of segments, at a rate chosen to
match that seen in network simulations, but qualitatively accurate
results are obtained with much less computational cost.
In the global defect case, the energy is
not localized into the string cores and their resolution is not
essential. This enables the calculation of \CMB contributions from global
defects using field theory simulations \cite{Pen:1997ae,Durrer:1998rw}.

None of these three approaches is useful for the study of semilocal
strings, since they are not well-modelled by Nambu--Goto strings
while, unlike the global case, the string cores are potentially
important. Fortunately, in a recent breakthrough we performed the
first \CMB calculations for local cosmic strings to derive from
field-theoretical simulations \cite{Bevis:2006mj}. We achieved this by the use of a parametrized
approximation to reduce the range of scales required in the simulations, but which had no resolvable effect on the \CMB predictions. That approach has opened the door for the possibility of
simulating defects with no known approximating models.

We have therefore employed the machinery developed for the local
string case to the simplest model containing semilocal strings, with
the parameters chosen to give the Bogomol'nyi--Prasad--Sommerfield
(BPS) limit (see Section \ref{sec:model}). The BPS limit arises naturally
in D-term SUSY inflation and $D3/D7$ brane inflation. We leave the
study of non-BPS cases to a future article but it is the expectation
that in the two opposite extremes between which the BPS case lies, the
semilocal result will tend to that of global textures or to that of
local strings
\cite{Achucarro:1997cx,Achucarro:1998ux,Achucarro:2005tu}. We perform
a detailed comparison of our results with these two cases using our
earlier results in Ref.~\cite{\longprd} for local strings. For the
case of global textures, we  perform \CMB
calculations using the linear $\sigma$ model.\footnote{
See  for instance Ref.~\cite{Durrer:1995sf} 
for a comparison of the dynamics of the linear and non-linear $\sigma$ models.}
 We leave a discussion of our texture calculations
to an Appendix.

We have also performed a multi-parameter comparison with current data,
yielding results similar to those of Ref.~\cite{Bevis:2007gh} where
it was seen that local cosmic strings are mildly favoured by current
\CMB data. We find that the preference for semilocal strings
is slightly greater and, as in Ref.~\cite{Bevis:2007gh}, we explore
this preference using Bayesian evidence calculations, as well as
considering the implications of non-\CMB data.

This paper is organized as follows: in the next Section we discuss
the semilocal string model, before giving a brief overview of the
simulation and \CMB calculation method in Section \ref{method}. We
will then present our \CMB results and consistency checks in
Section~\ref{CMBresults}, including the comparison with local strings and
global textures. Section \ref{MCMC} then presents the comparison with
data.

\section{Semilocal string model}
\label{sec:model}

The simplest model giving rise to semilocal strings is given by the
Lagrangian density \cite{Vachaspati:1991dz} 
\bea 
\mathcal{L}
&=&\left|D_{\mu}\phi_1\right|^2+\left|D_{\mu}\phi_2\right|^2-
\frac{1}{4 e^2} F_{\mu\nu} F^{\mu\nu}\nonumber\\ & & -
\frac{\lambda}{4}\left(\left|\phi_1\right|^2 +
\left|\phi_2\right|^2-\eta^2\right)^2  \,,
\label{lagr}
\eea 
where $D_\mu=\partial_\mu+iA_\mu$ and $F_{\mu\nu}=\partial_\mu A_\nu -
\partial_\nu A_\mu$. The fields $\phi_1$ and $\phi_2$ are complex
scalar fields and $A_\mu$ is an Abelian gauge field. The parameter $\eta$ 
gives the symmetry breaking scale. 
This Lagrangian possesses all the necessary ingredients to give rise
to semilocal 
strings and so allows insight into more complicated models, such as
D-term inflation in $N=1$ or $N=2$ supersymmetric (SUSY) models
\cite{Achucarro:2001ii,Pickles:2002ym} and $D3/D7$ brane inflation
\cite{Dasgupta:2004dw}.

Since this model is just the Abelian Higgs model with an extra Higgs
field, the semilocal strings arising from it can be seen as embedded
Nielsen--Olesen strings \cite{Achucarro:1999it}. The vacuum manifold
is $S^3$, which is simply connected, and the existence and stability
of the strings depend on energetic and dynamical causes, not on the
topology of the vacuum. There is only one parameter governing the
stability of the strings, namely, $\beta^2 \equiv \lambda/2e^2$: for
$\beta<1$ they are stable (even quantum mechanically); for $\beta>1$
they are unstable. The BPS case ($\beta=1$) is the transitional case,
where the strings are neutrally stable
\cite{Hindmarsh:1991jq,Hindmarsh:1992yy}. There is a zero mode
concerning the string width in the BPS case: all strings with widths
that range from the width of an Abelian Higgs string to infinity are
degenerate in energy. But even though the string can have a very large
(infinite) core, the magnetic flux is nevertheless quantized. It is
also worth noting that when two semilocal strings of any width
collide, they intercommute \cite{Leese:1992fn,Laguna:2006qr,Eto:2006db}
 and their widths may revert to 
the minimum width (i.e., the
width of an Abelian Higgs string) \cite{Laguna:2006qr}.
As mentioned earlier, the BPS case arises naturally in, for example, 
D-term inflationary scenarios,
In more generic cases, non-BPS cases will arise. We can anticipate
some qualitative behaviour for the non-BPS case. For $\beta>1$ the semilocal string become
unstable, and they will behave like  global textures. For $\beta<1$, the lower the value of $\beta$, 
the more 
semilocal
string networks resemble Abelian Higgs string networks \cite{Achucarro:2005tu}, and their CMB predictions
will tend to that of Abelian Higgs strings.

One important difference between Abelian Higgs cosmic strings and
semilocal strings is that the latter are non-topological, so they can
have ends. We therefore do not expect infinite semilocal strings to be
formed by the Kibble mechanism in a cosmological phase
transition. Instead, a collection of segments forms. The evolution of
the segments is very complicated: some will shrink to zero, others
will join both ends to form loops, while some will join to nearby
segments to make longer segments, and eventually form virtually
infinite strings \cite{Achucarro:2005tu}. At the BPS limit, the
formation of longer segments is strongly suppressed, which led the
authors of Refs.~\cite{Urrestilla:2004eh,Dasgupta:2004dw} to
conjecture that no strings would be formed in such a model after
inflation, and to claim that the D-term inflation without cosmic
strings could take place.

While the semilocal model can be reduced to the Abelian Higgs model by
removing one complex scalar field, it can also be reduced to a
global texture model by removing the gauge field (see Appendix). 
A key question is
how those removals modify the predicted anisotropy, keeping
fundamental Lagrangian parameters fixed. An important derived
parameter is the string mass per unit length in the Abelian Higgs and
semilocal cases \cite{Bogomolnyi:1976a,Achucarro:1999it}: 
\be
\label{mu}
\mu \equiv 2\pi \eta^2 \,,
\ee
and the \CMB anisotropies are proportional $(G\mu)^{2}$, where $G$ is
the gravitational constant. We continue to use this measure even when
talking about the texture model, but in that case it has no direct
interpretation and is merely a measure of the energy scale of the
model. It is known that textures give significantly lower
anisotropies than strings for a given $G\mu$, since textures decay
much more quickly once inside the horizon. Our original preconception
was that the anisotropies from semilocal strings would be of the same
order as, but a little smaller than, those of textures. This
expectation is because the gauge field cancels some of the field
gradients present in the texture case and contributes little to the
energy density \cite{Hindmarsh:1992yy}, while in the BPS limit there
are no long strings and hence no expectation of large anisotropies
from string-like sources. Indeed, we will find that this expectation
is confirmed by our detailed analysis.

\section{CMB calculation method}
\label{method}
\subsection{Overview}

As discussed in the introduction, the method used to calculate \CMB
power spectra employed here is exactly that used for Abelian Higgs
strings in Ref.~\cite{Bevis:2006mj} and we refer the interested reader
to that article for greater depth. However we will briefly describe
the fundamentals of the approach in this Section.

Since observations reveal that the cosmological perturbations are
small (except on short scales in the current epoch) the defects
can be taken to evolve in an expanding flat
Friedmann--Lema\^itre--Robertson--Walker universe. Inflation is taken
to set up small primordial fluctuations, which then evolve passively
under the Einstein--Boltzmann equations. However, the defects perturb
the metric, via their energy--momentum tensor, and these 
perturbations also have to be taken into account when computing \CMB
power spectra. But since the perturbations introduced by either
inflation or defects are very small, any coupling between them is
insignificant and the two perturbation sets may be calculated
separately. The result is two statistically independent sets
of \CMB anisotropies, the power spectra of which simply add to give
the total.

Fundamentally the \CMB power spectra are two-point correlation
functions of the \CMB temperature (or polarization) fluctuations, and
in the defect case they can be determined from the two-point unequal-time
correlation functions (\textsc{uetc}s) of the defect energy--momentum tensor
$\FT{T}_{\mu\nu}$
\cite{Turok:1996ud,Pen:1997ae,Durrer:1997ep,Durrer:2001cg}: 
\be
\FT{U}_{\kappa\lambda\mu\nu}(\vect{k},\tau,\tau') = \left<
\FT{T}_{\kappa\lambda}(\vect{k},\tau)
\FT{T}_{\mu\nu}\conj(\vect{k},\tau') \right>.
\label{uetc}
\ee
Here $\vect{k}$ is the comoving wavevector, $\tau$ and $\tau'$ are
two conformal times and $\sim$ denotes a Fourier space quantity, as in the notation of Ref.~\cite{\longprd}.

However, the above 55 complex functions of 3 variables may be reduced to just 5 real functions of two variables, via
the use of statistical isotropy, energy--momentum conservation, and a
property of defect networks known as \emph{scaling}. The last property
derives from the causal nature of the network: that field smoothing is
limited by the horizon and, as such, defect networks often tend to an
attractor regime in which statistical measures of their distribution
are a function of a single quantity: the horizon size $\tau$. Assuming
scaling and statistical isotropy, the \UETCs can be written as
\cite{Durrer:2001cg,\longprd} 
\begin{eqnarray}
 \FT{U}_{\kappa\lambda\mu\nu}(k,\tau,\tau)
 & = &
 \frac{\eta^4}{\sqrt{\tau\,\tau'}} \frac{1}{V}
 \;
 \FT{C}_{\kappa\lambda\mu\nu}(k\tau,k\tau')
\label{uetc5}
\end{eqnarray}
where $V$ is the fiducial simulation volume and $\FT{C}$ is the
scaling function of the \textsc{uetc}. Further, the functions $\FT{C}$ decay
for large or small $\tau/\tau'$ or for large $k\sqrt{\tau\tau'}$ and
hence their measurement is only required over a fairly limited
parameter space.  

Of the 5 independent scaling functions, 3 of them represent scalar
degrees of freedom ($\FT{C}\Sr_{11}(k\tau,k\tau')$,
$\FT{C}\Sr_{12}(k\tau,k\tau')$,$\FT{C}\Sr_{22}(k\tau,k\tau')$) while
the remaining two represent the vector and tensor degrees of freedom
respectively ($\FT{C}\Vr(k\tau,k\tau')$, $\FT{C}\Tr(k\tau,k\tau')$)
\cite{\longprd,Durrer:1997ep}. Unlike the inflationary case, the
amplitude of the tensor perturbations is fixed and directly obtained
from the simulations, and it is not possible to neglect the vector
perturbations as they are continuously sourced by the presence of the
defect network.

The \UETC scaling functions are then fed into a modified version of
the \CMBEASY Boltzmann code \cite{CMBEASY} which computes the \CMB
temperature and polarization power spectra. However, in order for \CMBEASY to use
the \UETC data, the functions $\FT{C}(k\tau,k\tau')$ must be first decomposed as:
\be 
 \FT{C}(k\tau,k\tau')
 = \sum_{n} \lambda_{n}
    \FT{c}_{n}(k\tau) \;
    \FT{c}_{n}(k\tau'),
\label{eigens}
\ee 
which in the numerical case is tantamount to determining the
eigenvalues and eigenvectors of a real, symmetric matrix. Note that
$\FT{C}\Sr_{12}$ is not symmetric, but 
$\FT{C}\Sr_{21}(k\tau,k\tau')=\FT{C}\Sr_{12}(k\tau',k\tau)$ and so the $M\times M$ matrices
representing the scalar degress of freedom ($\FT{C}\Sr_{11}$,$\FT{C}\Sr_{12}$,$\FT{C}\Sr_{21}$ and
$\FT{C}\Sr_{22}$) are tiled together to yield a symmetric $2M\times 2M$ matrix  which 
is then decomposed this way.

The modified \CMBEASY then takes a single
eigenvector $\FT{c}_{n}$ as a source function and determines the
corresponding contribution to the \CMB power spectra, with summation
over all such contributions yielding the total. We found that for the
semilocal case, the number of eigenvectors that should be included in
the computation of the power spectra to achieve convergence was lower
than that needed in the Abelian Higgs case; nevertheless, we use the
same number of eigenvectors as in the Abelian Higgs case (see Table
\ref{parameters}) to minimize the differences in the computation. Also
there are some choices about how to decompose the \UETCs into the form
of Eq.~(\ref{eigens}), and again we follow exactly the method of
Ref.~\cite{\longprd}.  

Note that scaling is broken at the radiation--matter
transition and therefore the scaling \UETC functions must be
determined for the defects under both matter and radiation-domination. These are decomposed into eigenvectors and fed into the
modified \CMBEASY code, which then interpolates between the radiation
and matter results to model the transition. For the transition
into the $\Lambda$--dominated epoch, the sources simply decay 
and simulation is not
required. In any case, perturbations sourced at such late times
 affect only the very largest scales.

\subsection{Semilocal simulations}

In order to calculate the scaling \UETC functions for the semilocal
string model above, we perform field theory simulations under both
radiation- and matter-domination. However, since at times of importance
for \CMB calculations the string width is many orders of magnitude
smaller than the horizon size, it is not possible to perform simulations
for such times while still resolving the string cores. Fortunately,
scaling implies that this is not required and that the scaling \UETC
functions may be obtained from simulations of early times, with
$\tau\sim100\eta^{-1}$, and then used to describe the late-time
behaviour of the system. Note that although we perform simulations in a matter-dominated background
for $\tau \sim 100\eta^{-1}$, when the real universe was
radiation-dominated, scaling implies that the network properties are the same
at $100\eta^{-1}$ as they are in the true late-time matter era.

However, performing the simulations required with current or near
future computational technology is a considerable challenge. The
(minimum) string width ($\rmin$) is fixed in physical coordinates and, 
in the comoving coordinates required for the simulations, it varies as:
\be
\rmin=\frac{1}{a\sqrt{\lambda}\eta}=\frac{1}{2ae\eta}\,,
\label{rmin}
\ee
where $a$ is the cosmic scale factor. It therefore shrinks 
as $\tau^{-1}$ under a radiation-  or as
$\tau^{-2}$ under a matter-dominated universe. If the strings are resolved at
the end of the simulation, then they are enormous and larger than the horizon size for times only a
little earlier, particularly under matter domination. The system will not scale, or even contain significant regions where the fields
lie on the vacuum manifold, if $\tau$ is comparable to $\rmin$. Thus, the \UETC data
 can only be collected over a limited region of the required parameter space.

A simple means of circumventing the above problem is to 
promote the parameters $\lambda$ and $e$ in the action to
time-dependent variables \cite{\longprd} as:
\bea
\lambda&=&\lambda_0\;a^{-2(1-s)}\nonumber\\
e&=&e_0\;a^{-(1-s)}.
\label{eqns}
\eea
Now the scale $\rmin$ varies as:
\be
\rmin =  r_{\mathrm{min},0}\;a^{-s},
\ee
and variation of the action yields equations of motion as:
\bea
&&\ddot{\phi_1} + 2 \frac{\dot{a}}{a} \dot{\phi_1} - D_{j} D_{j} \phi_1 
  = 
 -a^{2s} \frac{\lambda_{0}}{2} \left( |\phi_1|^{2}+|\phi_2|^{2} -
\eta^{2} \right)\!\phi_1 \nonumber\\ 
&&\ddot{\phi_2} + 2 \frac{\dot{a}}{a} \dot{\phi_2} - D_{j} D_{j} \phi_2 
  =  
 -a^{2s} \frac{\lambda_{0}}{2} \left( |\phi_1|^{2}+|\phi_2|^{2} -
\eta^{2} \right)\!\phi_2 \nonumber\\ 
&&\dot{F}_{\!0j} + 2 (1\!-\!s) \frac{\dot{a}}{a} F_{\!0j} -
\partial_{i}F_{\!ij} 
  =\nonumber\\
&&\qquad \qquad -2a^{2s} e_{0}^{2} \; \mathcal{I}m \! \left
[ \phi_1\conj D_{j}\phi_1 +   \phi_2\conj D_{j}\phi_2\right].
 \label{eom}
\eea
For the case of $s=1$, the true dynamics are obtained, while for $s=0$
the factors of $a$ responsible for the shrinkage of the (minimum) string width 
are removed from the equations of motion and $\rmin$ becomes comoving. This is similar to methods of Refs. \cite{Press:1989id,Moore:2001px}, where the undesirable factors of $a$ were directly removed from the dynamical equations, but here the parameter $s$ can be varied and its effects measured. This is particularly important because $s\ne1$ means that energy--momentum conservation is violated and it is $T_{\mu\nu}$ that sources the perturbations. For the simulations under
radiation-domination we can actually simulate the $s=1$ case directly and  compare in detail our intermediate results for $s<1$ with the true case. However, one of the results of Ref.~\cite{\longprd} was that the
\CMB results for Abelian Higgs strings are insensitive to $s$, and here we extend that result to semilocal strings. 

There is good evidence from previous simulations of both Abelian Higgs and semilocal strings 
that the network results are independent of the initial conditions
once scaling is reached \cite{Vincent:1997cx,Achucarro:1998ux,Moore:2001px,Urrestilla:2001dd,\longprd}.
The goal is therefore to produce an initial configuration in a
numerically feasible way that obeys the analogue of Gauss' law in the
model and that will yield scaling fairly fast, in order to maximize
the dynamical range of the simulation. Following again
Ref.~\cite{\longprd} we set all temporal derivatives and gauge fields
to zero, and the fields $\phi_1$ and $\phi_2$ are set in such a way
that they lie in the vacuum manifold $\left( |\phi_1|^{2}+|\phi_2|^{2}
= \eta^{2} \right)$ with random phases in $S^3$.  The system is then
evolved using the equations of motion obtained after discretizing the
action via the standard Moriarty et al.~\cite{Moriarty:1988fx} scheme
and encoded in C++ using the LATfield Library \cite{LATfield}. This
scheme preserves the Gauss constraint during the evolution, even in
the case of $s<1$, because our equations are derived from the discretized
action (see Ref.~\cite{\longprd}).

\section{Simulation and \CMB power spectrum results}
\label{CMBresults}
\subsection{Numerical aspects and tests for scaling}

\begin{figure}[t]
\begin{center}
\includegraphics[width=8cm]{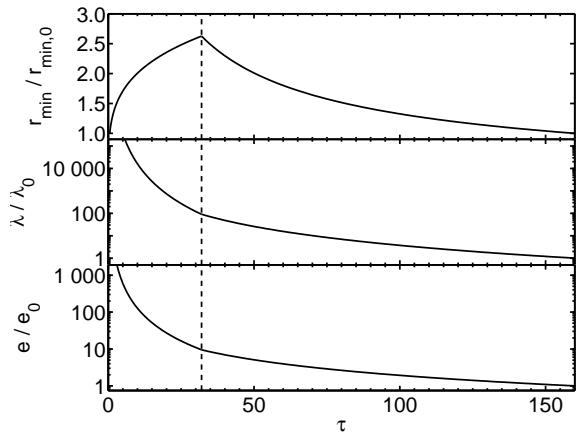}
\caption{\label{paramvary} Variation of $\rmin$ and  the parameters $e$ and
$\lambda$ for the $s=0.3$  simulations,  which mimic the one used in
Ref.~\cite{\longprd}. The subscript 0 denotes the value at the end of the simulation.}
\end{center}
\end{figure}

The numerical simulations were performed on the UK National Cosmology
Supercomputer \cite{cosmos-website}, where we simulated five different
realizations for both matter- and radiation-dominated epochs and for
$s=0.0$ and $s=0.3$, as well as $s=1$ for radiation alone.
Of those, only the
simulations with $s=0.3$ were used to produce the \CMB power spectra;
the others were used to test the validity of our approach.
We begin and end the simulation
with the same value for $\rmin$; that is, we
initially set $s$ to a negative value so that
$\rmin$ increases until a time $\tau_s$, after
which $s$ is positive so that $\rmin$
shrinks
during the primary part of the simulation (see Fig.\ref{paramvary}). This process accelerates
the formation of vacuum regions.

\begin{table}
\begin{ruledtabular}
\begin{tabular}{lc}
Parameter			&	Value\\
\hline
Lattice size $N$			&	512\\
Lattice spacing	$\Delta x$ $[\eta^{-1}]$		&	0.5\\
Time-step	$\Delta t$	$[\eta^{-1}]$		&	0.1\\
Scalar coupling $\lambda_0$		&	2.0\\
Gauge coupling $e_0$		&	1.0\\
Initial conformal time	 $\tau_{\rm i}$	$[\eta^{-1}]$&	1.0\\
Final conformal time $\tau_{\rm end}$	$[\eta^{-1}]$	&	128\\
Initial $s$					&	$-0.116$ \\
Final $s$					&	0.3\\
Time of change in s value	$[\eta^{-1}]$		&	32\\
\hline
$\tau_{\rm sim}$ range of $\xi$ fit	$[\eta^{-1}]$	&	50 - 128\\
Dynamic range $(\tau/\tau')_{\rm max}$  			&	$\sim 2$\\
\UETC sample matrix size			&	512\\
\UETC matrix sampling			&	linear\\
No.\ eigenvectors			&	128\\
\end{tabular}
\end{ruledtabular}
\caption{\label{parameters} Parameters used in production runs.  The
first part of the table lists the parameters of the simulations, with
the second part listing parameters of the \UETC method for calculating
the \CMB power spectrum. Note that the values are identical to those
of Ref.~\cite{\longprd}, except for the time range of the simulations. }
\end{table}

Table~\ref{parameters} shows the values of
the parameters used for the production runs. Most of the parameters
match those of Ref.~\cite{\longprd}; furthermore, the coupling
parameters were also varied as shown in Fig.~\ref{paramvary} to
replicate the behaviour of Ref.~\cite{\longprd}. Because of the
existence of zero modes in the semilocal system it is dangerous to run the
simulation for longer than the half-crossing-time. In order to keep
the dynamic range of Ref.~\cite{\longprd}, here we had to begin
taking \UETC data at an earlier time ($\tau=50$ rather than $\tau=64$). One might wonder whether
the system is already into the scaling regime by those early times,
but we will show in this Section that in fact this is the case.

One means to test for scaling is via the string length density, which should be proportional
to $\tau^{-2}$, since the average length of string per horizon volume should be proportional to
$\tau$ and the horizon volume varies as $\tau^{3}$. The detection of (topological) Abelian Higgs strings in a lattice
field simulation is fairly straightforward, since one can track, for
instance, zeros of the Higgs field or the winding 
of its phase around them. Alternatively one could use the average 
Lagrangian density (as in Ref.~\cite{\longprd}) or the $T_{00}$
component of the energy--momentum tensor, which should also vary as $\tau^{-2}$.

\begin{figure}[t]
\begin{center}
\includegraphics[width=8cm]{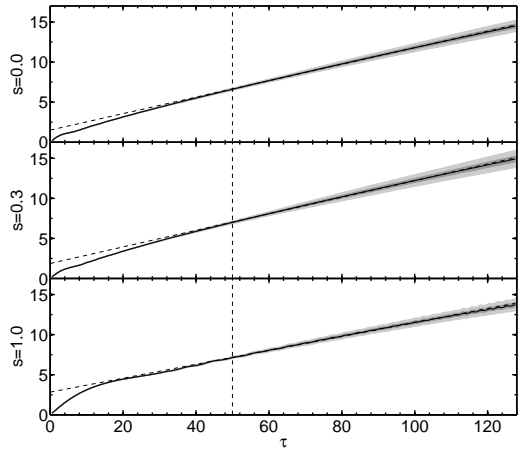}
\vspace{0cm}\caption{\label{xi} Average value of $\xi$, as defined in
Eq.~(\ref{defxi}), for $s=0.0,\,0.3,\,1.0$, in the radiation era. The
average is over five different realizations for each value of $s$, and
the shaded regions correspond to 1-$\sigma$ and 2-$\sigma$
deviations. The best-fit line for times $\tau>50$ is also shown (dashed line). Note
that the best-fit line is an excellent approximation, and that the
differences between different values of $s$ are minimal.  }
\end{center}
\end{figure}

In fact, for (non-topological) semilocal strings those alternative
methods must be employed: the zeros of the Higgs fields are not a good
measure, since there is no need for the string core to be a zero of
the Higgs field. Besides, semilocal strings in the Bogomol'nyi limit
($\beta=1$) are expected to form short segments that either collapse
very quickly or become fatter, diluting away the core of the string. That
is why we used a different measure to quantify the scaling of the semilocal
BPS string system.

\begin{figure*}[t]
\begin{center}
\includegraphics[width=5.8cm]{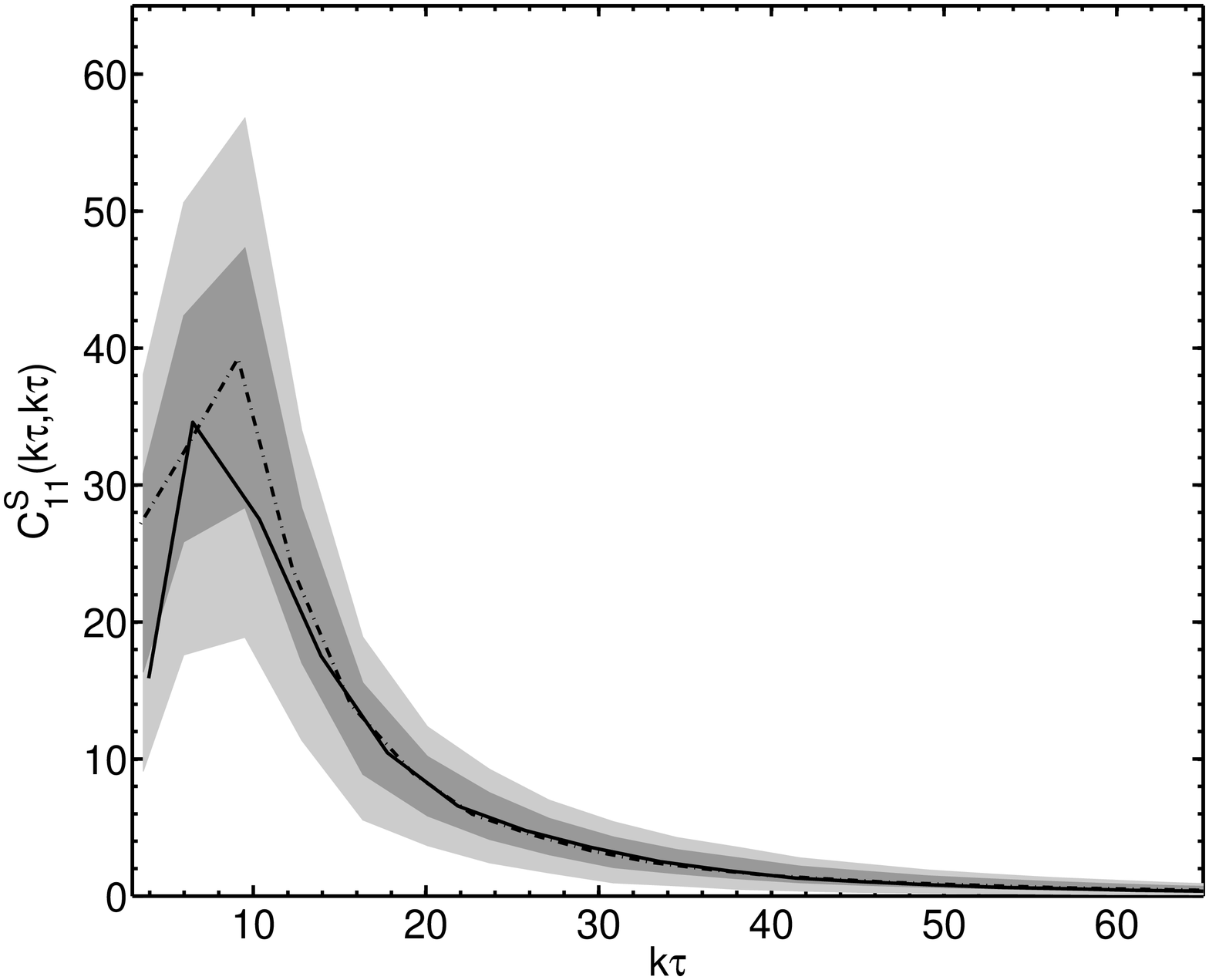}
\includegraphics[width=5.8cm]{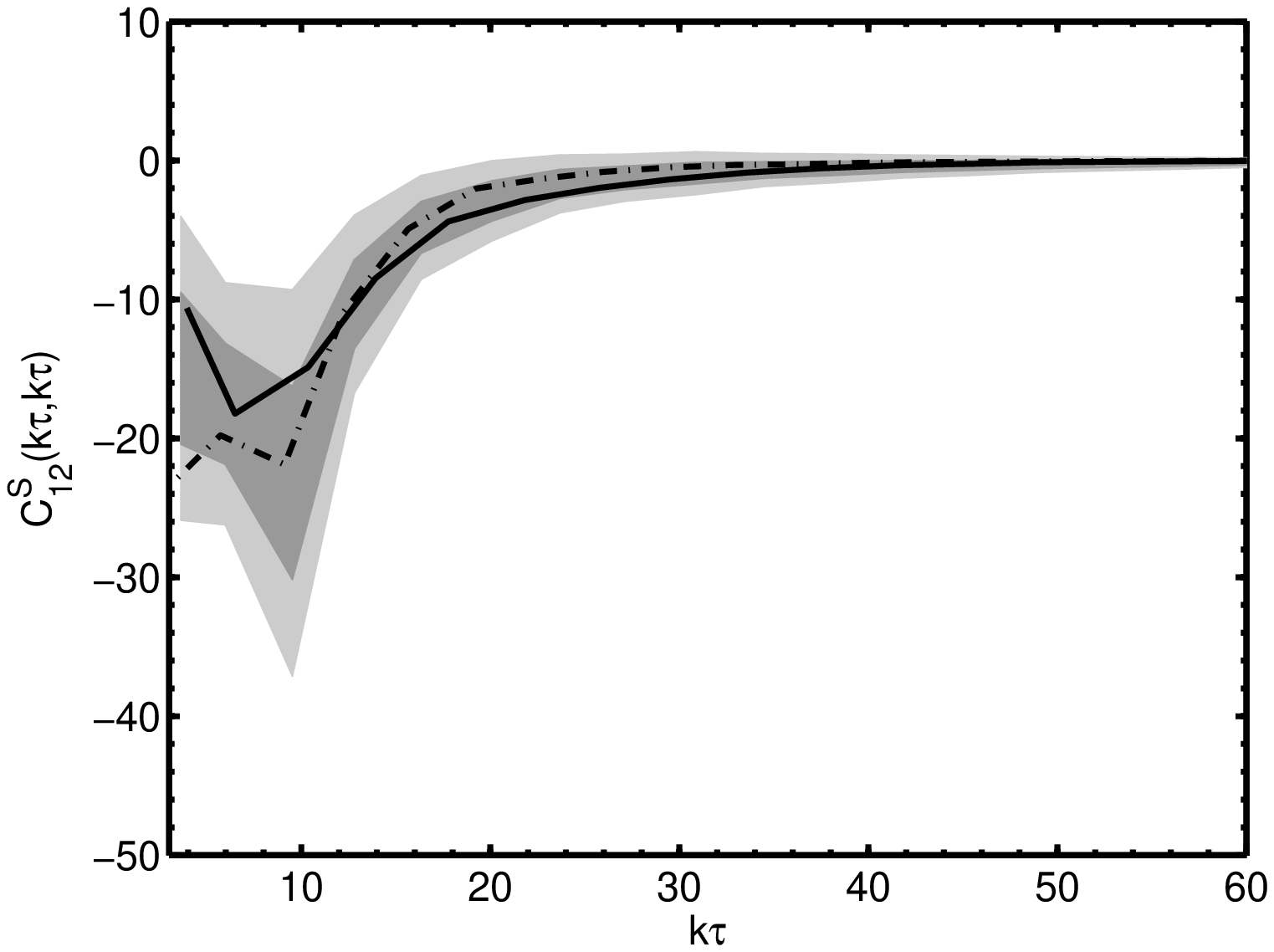}
\includegraphics[width=5.8cm]{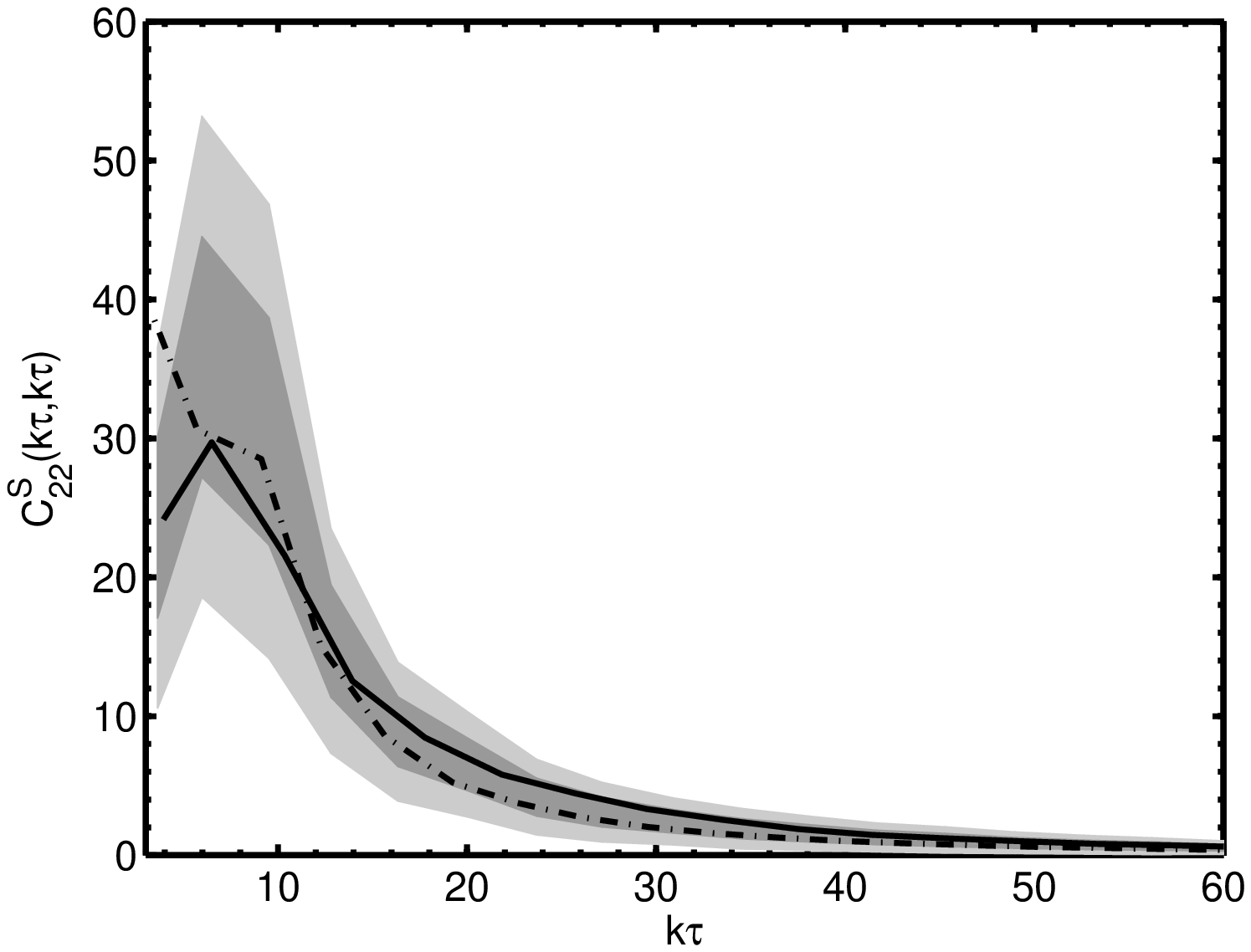}\\
\includegraphics[width=5.8cm]{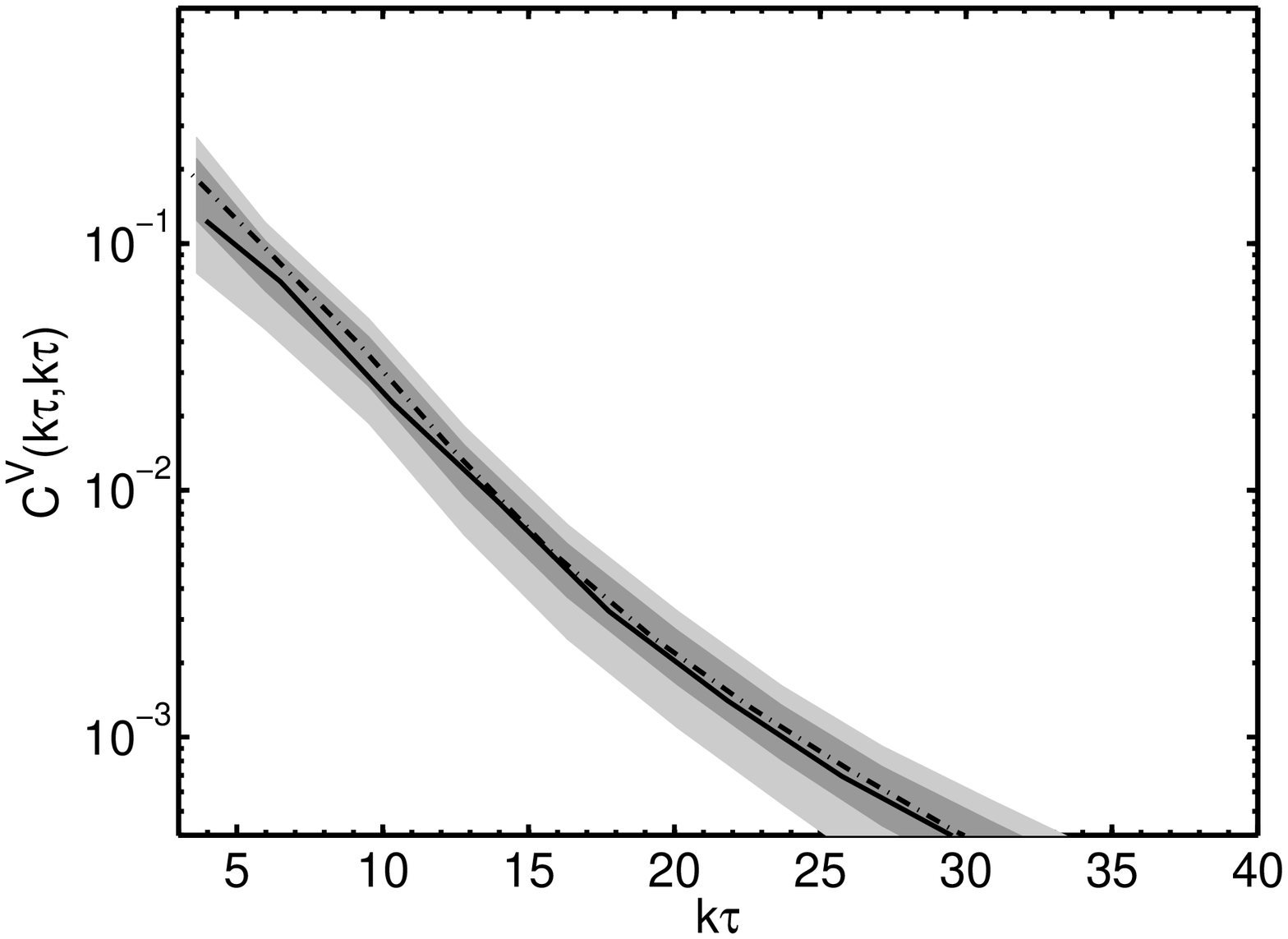}
\includegraphics[width=5.8cm]{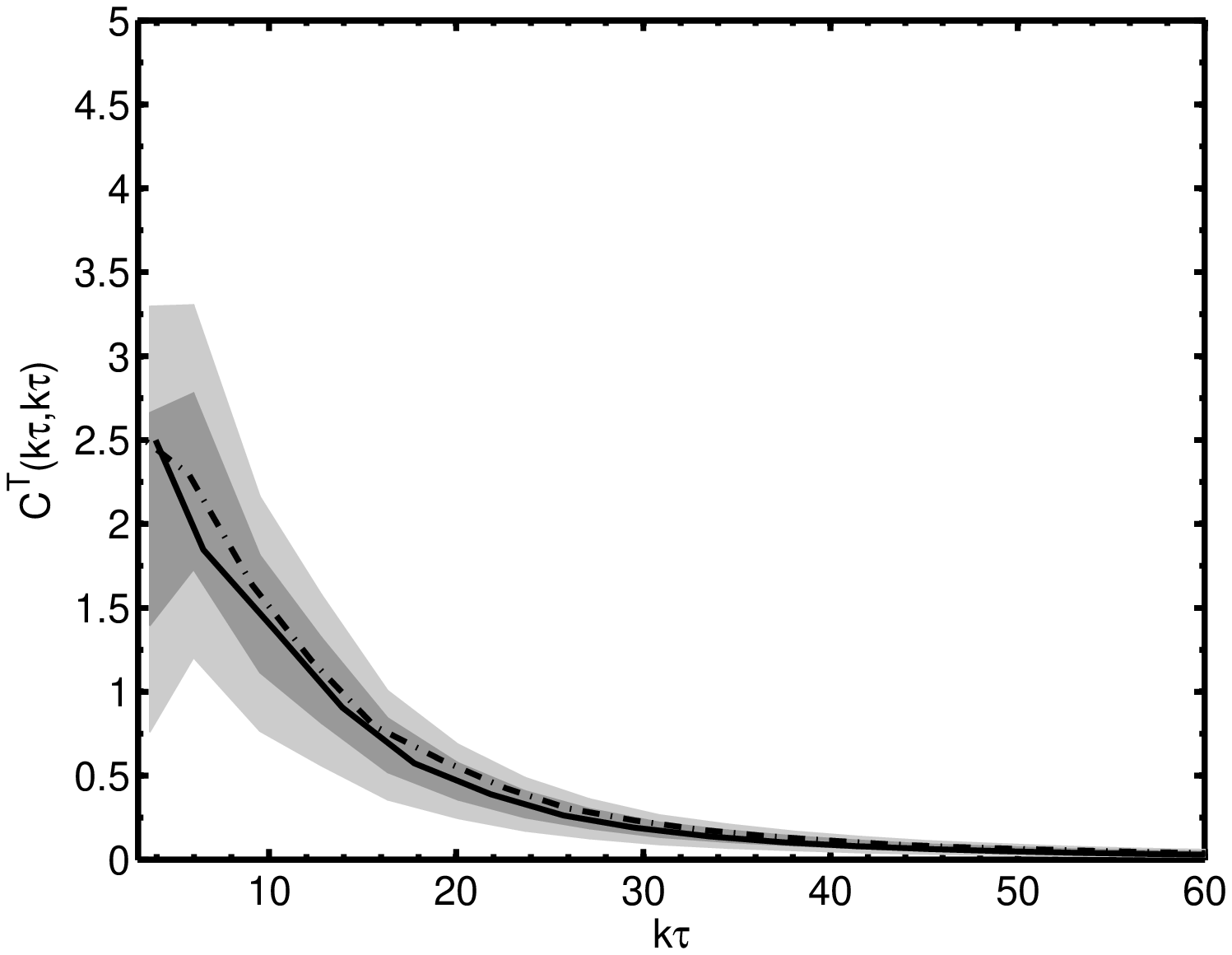}
\caption{\label{ETCscales} The dependence of the 5 
\ETCs upon $s$ in the radiation era at
$\tau_{\rm sim}=128\eta^{-1}$. The shaded regions show the 1-$\sigma$
and 2-$\sigma$ uncertainties for the $s=0.3$ case, while the lines
indicate the $s=1.0$ (solid), $s=0.0$ (dot-dash) results, which
clearly lie within the statistical uncertainties.}
\end{center}
\end{figure*}

In this work, we use the $T_{00}$ component of the
energy--momentum tensor to define our length measure $\xi$ as:
\be
\xi=\frac{\eta}{\sqrt{T_{00}}},
\label{defxi}
\ee
and use it to check the scaling of the system. Figure~\ref{xi} shows
the result of the simulations for the radiation era for
$s=0.0,\,0.3,\,1.0$, as well as the best fit for times $\tau>50$. First of
all, note that the differences with respect to varying the value of
$s$ are minimal. It is also clear from these figures that after the
initial non-scaling period, $\xi$ follows a linear behaviour
\be
 \xi \; \propto \; (\tau - \tau_{\xi=0}).
 \label{xioffset}
\ee
There is nothing fundamental about the offset $\tau_{\xi=0}$, as
argued for Abelian Higgs case in Ref.~\cite{\longprd}; it is a
consequence of the non-scaling initial period of the simulations and is negligible in the late-time 
limit of interest for \CMB calculations. What is fundamental is the value of
the slope of $\xi$, which shows little dependence upon $s$: $0.10\pm0.01$ for $s=0.0$, $0.10\pm0.02$ for $s=0.3$ and $0.09\pm0.02$ for $s=1.0$.
The difference between the three plots in Figure \ref{xi} is due almost entirely to differing values of $\tau_{\xi=0}$.

\begin{figure}[t]
\begin{center}
\includegraphics[width=7cm]{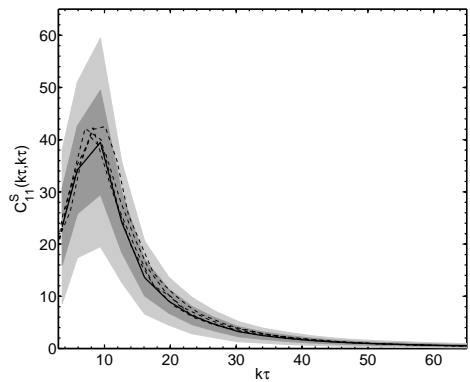}
\vspace{0cm}\caption{\label{ETCscale} The equal-time scaling function
$\FT{C}\Sr_{11}(k\tau,k\tau)$ averaged over five realizations for
$s=0.3$ in the radiation era. The different lines correspond to
roughly uniformly spaced $\tau$ values in the range
$50\eta^{-1}<\tau<128\eta^{-1}$. The shaded regions show the
1-$\sigma$ and 2-$\sigma$ uncertainties in the mean indicated for the
latest time (solid). For visualization purposes, the mean offset across the five
realizations is used, whereas the actual
\CMB calculations use independent offsets for each realization.}
\end{center}
\end{figure}

However, given that the system scales not with $\tau$ but with $\tau-\tau_{\xi=0}$, then following Ref.~\cite{\longprd} we perform a rescaling of the measured \UETC functions $\FT{C}(k\tau,k\tau')$ in accordance with:
\be
\tau\rightarrow\tau-\tau_{\xi=0}.
\ee
These functions then provide a more important test of scaling for the present work:
 after the rescaling, do they show negligible absolute time dependence? Figures \ref{ETCscales} and \ref{ETCscale} show the 
equal-time correlator (\textsc{etc}) scaling functions, that is when $\tau=\tau'$. These are 
the most important part of the \textsc{uetc}s, and the most suitable for visualization. The
different lines in Fig.~\ref{ETCscale}
correspond to the \ETC function averaged
over five realizations, for different times; the shaded regions
correspond to the 1-$\sigma$ and 2-$\sigma$ uncertainties in the mean
at the last time step (when the system is closest to scaling). 
Clearly, the
different lines lie within the statistical uncertainties of the
simulation, and there is no absolute dependence of the \ETC on
time. As the \UETCs are the only quantities entering the \CMB power
spectrum calculations, we are confident that the earlier sampling does
not affect the results.

We can also make use of the \ETCs to show that the dependence of the
simulations with $s$ is quite mild. Figure~\ref{ETCscales} shows the
\ETCs for the last time step in the simulation for $s=0.0,\, 0.3,\,
1.0$ for the five different \UETCs need for  the  \CMB
calculation. In all cases the dependence on $s$ lies well within the
statistical uncertainties.

\begin{figure}[t]
\begin{center}
\includegraphics[width=8cm]{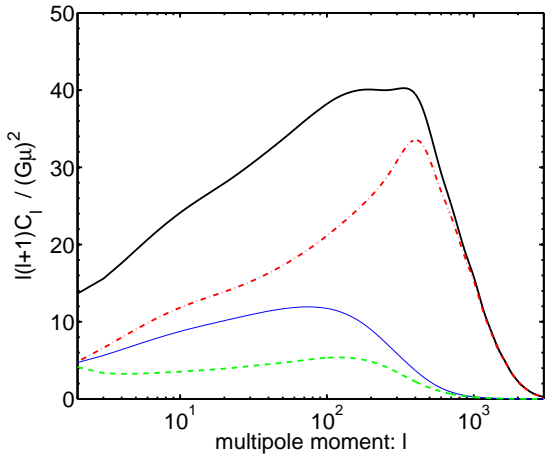}
\vspace{0cm}\caption{\label{SLSVT} \CMB temperature power spectrum
from semilocal string simulations. The solid black line is the total
prediction, which is obtained by adding the contribution of scalar
(dash-dot red), vector (thin blue) and tensor (dashed green) modes.}
\end{center}
\end{figure}

\subsection{Temperature power spectra from semilocal strings}

After checking that our numerical approach for simulating semilocal
strings is satisfactory by monitoring the influence of the parameter
$s$ in the simulations and confirming the early onset of scaling, the
resulting \UETCs were fed into the modified \CMBEASY Boltzmann code as explained
in Section~\ref{method}.  The perturbations obtained from the strings
and from inflation are calculated separately and added in quadrature
-- the tiny interaction between the two contributions can be safely
neglected. Besides, it is numerically quite slow to recompute the
string spectra for many different values of the cosmological
parameters. For these reasons, the string spectra were computed for a
fixed value of the cosmological parameters (the central values given
by cosmological experiments\footnote{
\label{parms} The calculation of the string spectra was done for
$h\!=\!0.72$ \cite{Freedman:2000cf}, $\Obhh\!=\!0.0214$
\cite{Kirkman:2003uv}, $\Ol\!=\! 0.75$
\cite{Knop:2003iy}.  
We also assumed spatial flatness and the optical
depth to the last-scattering surface $\tau=0.1$.}). We then tested at
the end that the string spectra do not change significantly when the
cosmological parameters are varied within the limits imposed by the
\CMB data. Note that this is the only point where the cosmological parameters
are involved in the calculation of the \CMB spectra.

\begin{figure}[t]
\begin{center}
\includegraphics[width=8cm]{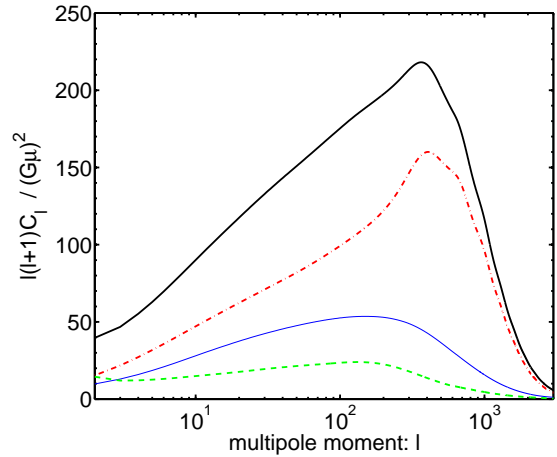}\\
\includegraphics[width=8cm]{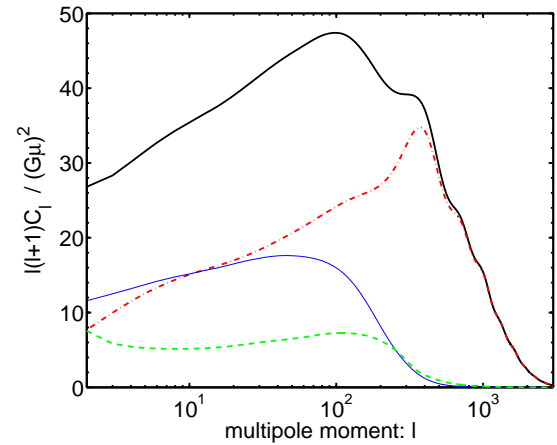}
\caption{\label{SVT} \CMB temperature power spectrum from Abelian Higgs strings (left) and texture (right) simulations. The solid black
line is the total prediction, and its constituents: scalar (dash-dot
red), vector (thin blue) and tensor (dashed green). Note the
difference in the scale between the two graphs.}
\end{center}
\end{figure}

Figure \ref{SLSVT} shows the \CMB temperature power spectrum from semilocal
cosmic strings, divided into scalar, vector and tensor components.
For comparison, we plot the corresponding power spectra from Abelian Higgs
strings and textures in Fig.~\ref{SVT}. The power spectra for Abelian Higgs
strings are taken from Ref.~\cite{\longprd},\footnote{Refs.~\cite{\longprd,Bevis:2007qz}
 employed a code in which a bug has been discovered, and this had a
 small effect in Ref.~\cite{Bevis:2007gh} since it used
their results
 directly (see the respective errata). Here we have used the corrected power
spectra from Refs.~\cite{\longprd,Bevis:2007qz} and quote the corrected results
 from Ref.~\cite{Bevis:2007gh}.}
 whereas the texture
spectra have been calculated by using new simulations of the texture
linear-$\sigma$ model obtained by setting to zero the gauge fields in
the semilocal model Eq.~(\ref{lagr}) (see Appendix for details on the
texture calculations). 

\begin{figure}[t]
\begin{center}
\includegraphics[width=9cm]{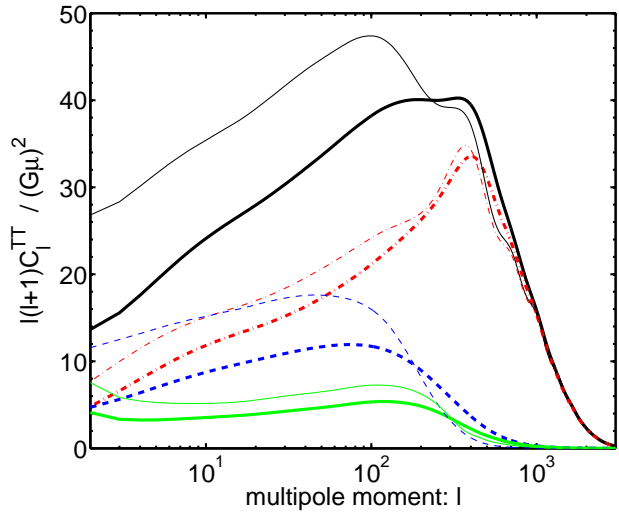}
\vspace{0cm} \caption{\label{sltex}The temperature power spectra from
semilocal strings (thick) and textures (thin), showing the scalar
(dash-dot red), vector (dashed blue) and tensor (lower solid green)
components. }
\end{center}
\end{figure}

The first thing to notice from these graphs is that, at fixed $G\mu$,
the Abelian Higgs strings give significantly higher anisotropies than
the other two by a factor of about 5. Semilocal strings and textures
are quite similar, with the former being slightly lower. This is in
good accord with the intuition that semilocal strings ought to be
similar to textures, but with some of the anisotropies cancelled by
the gauge field. Ultimately, this difference in anisotropy amplitude
will result in a weakening of observational constraints on $G\mu$ for
semilocal and texture as compared to Abelian Higgs strings.

Going beyond the amplitude, it is clear that the semilocal temperature
power spectra shape shares properties with both the Abelian Higgs and
the texture prediction. It is close to the texture prediction in, e.g., 
the level of the power of the anisotropies at fixed $G\mu$, 
although it does not have the small oscillations. On the other hand,
it peaks for larger values of $\ell$ than textures, more like
Abelian Higgs strings.

\begin{table}
\begin{ruledtabular}
\begin{tabular}{ll}
&	$G\mu_{10}$\\
\hline
Abelian Higgs string &	$2.7\times 10^{-6}$\\
Semilocal string &	$5.3\times 10^{-6}$\\
Textures  &	$4.5\times 10^{-6}$\\
\end{tabular}
\end{ruledtabular}
\caption{\label{gmu10} Normalization of different defects to match the
observed $\ell=10$ multipole value. }
\end{table}

\begin{figure}[t]
\begin{center}
\includegraphics[width=8cm]{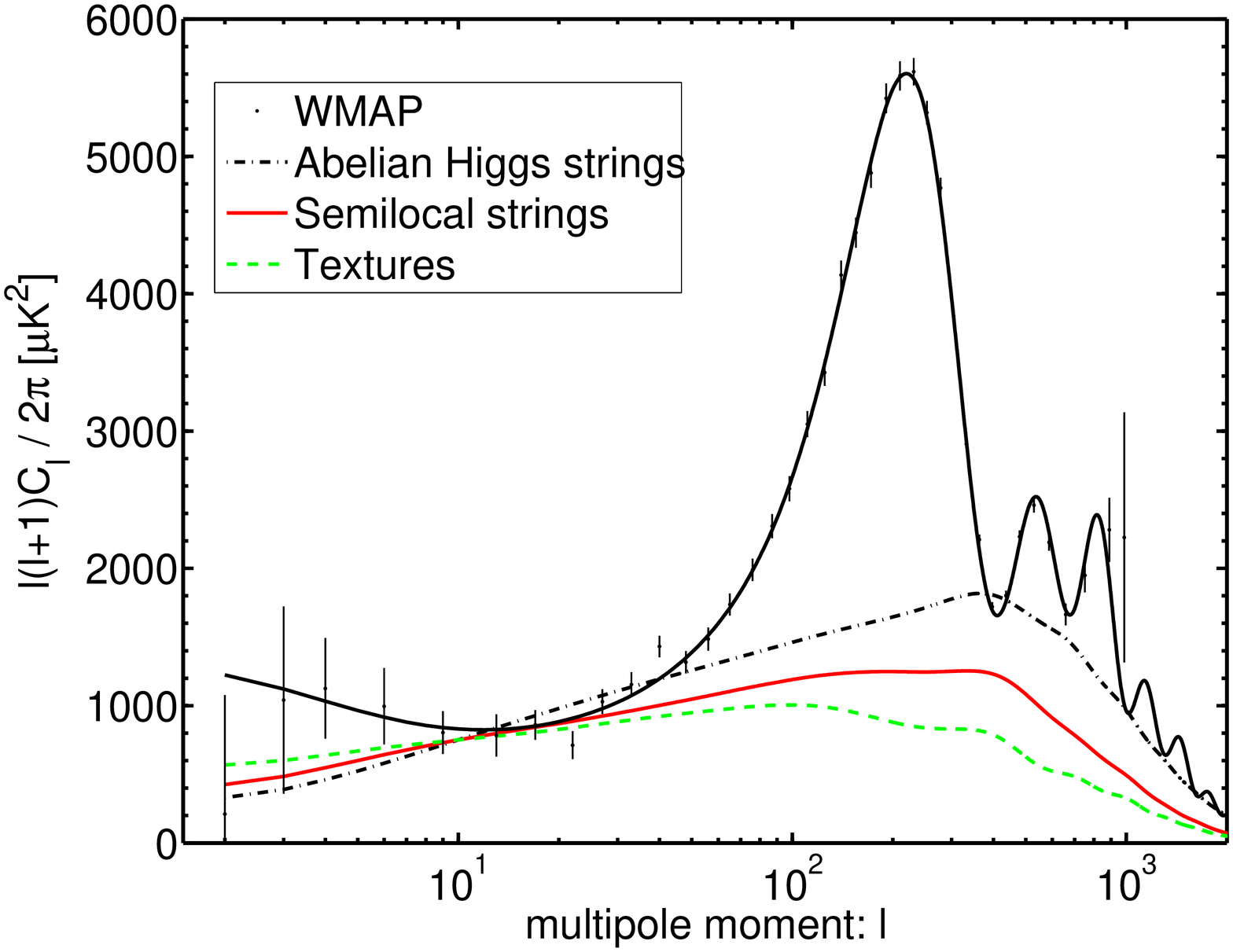}
\vspace{0cm}\caption{\label{gmu10fig} The temperature power spectra
for Abelian Higgs strings, semilocal strings, textures and the best fit
inflationary model. The points represent the three-year \WMAP
experimental data \cite{Hinshaw:2006ia}. The curves for defects 
are normalized to match the data at $\ell=10$.}
\end{center}
\end{figure}

Figure \ref{sltex} shows a direct comparison between the semilocal
strings prediction and the texture prediction for the temperature power
spectrum. As discussed before, the texture model can be obtained by
setting to zero the gauge coupling constant in the semilocal model, so
the differences appearing in the curves arise from coupling the scalar
fields to the gauge fields.  The overall shape of the curves for each
model is roughly the same, with that for textures being somewhat higher and
peaking earlier than that for the semilocal strings. If the Abelian Higgs model
prediction was plotted in the same graph, it would lie way above this
curve, since the overall scale of the spectrum is much higher.

However, bear in mind that the normalization of the spectra is a free
parameter, and one would like to compare the relative shapes of the
predictions to understand their detailed observational implications.
For example, it is customary to encode the information about the
overall power coming from the defects by $G\mu_{10}$.  The
normalization of the strings is proportional to $(G\mu)^2$.  Thus,
$G\mu_{10}$ is the value of the normalization by which the defect
spectra matches the \WMAP data \cite{Hinshaw:2006ia}
at multipole $\ell=10$, shown in
Table~\ref{gmu10} for each type of defect.  This is merely a tool to
compare different predictions.  Figure~\ref{gmu10fig} shows
pictorially the different defect predictions with these
normalizations, together with the experimental data and the
inflationary best-fit model.  Note that for textures it is not natural
to talk about ``energy per unit length'', but we adopt the same
normalization strategy Eq.~(\ref{mu}) by analogy with the strings, in
order to ease the comparison between models. Different normalization
schemes are discussed in the Appendix.

\subsection{Polarization power spectra from semilocal strings}

\begin{figure}[t]
\begin{center}
\includegraphics[width=9cm]{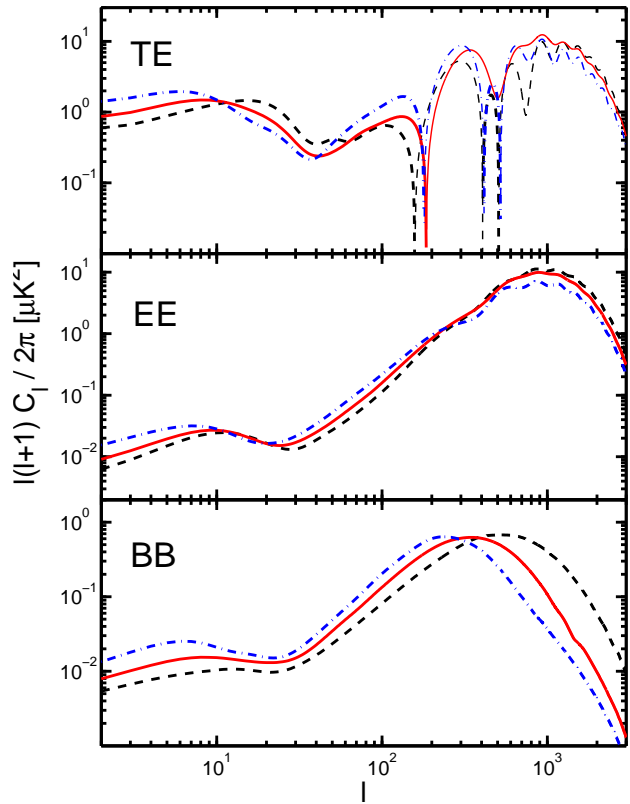}
\caption{\label{polar} Polarization power spectra for semilocal
strings (solid red), compared to Abelian Higgs strings (dashed black)
and textures (dot-dash blue). The top figure is the \TE power spectra,
the \EE is in the middle, and \BB at the bottom. All the curves are
normalized by making the temperature spectra match the \WMAP $\ell=10$
value, i.e., using the values of $G\mu=G\mu_{10}$ as in table
\ref{gmu10}.  In the \TE figure, the thick lines correspond to
correlations and the thin lines to anticorrelations.  The results
plotted here correspond to a value $\tau=0.1$ (see footnote
\ref{parms}).}
\end{center} 
\end{figure}

The polarization power spectra are shown in Fig.~\ref{polar},
decomposed as usual into the \EE, \TE and \BB components
\cite{Hu:1997hp}. The curves are normalized so they have equal \TT 
power at $\ell=10$ (and match the \WMAP measurement on this
scale).\footnote{ See Fig.~\ref{pol2} in Section \ref{MCMC} for
a figure with the normalization of each case set to match the respective 95\%
upper limits from current data.} 
Note that, as in Ref.~\cite{Bevis:2007qz}, we do not include the coupling
between inflation and defect perturbations as a result of
gravitational lensing at late times. We do not, however, expect this
effect to be significant: the gravitational lensing of the inflationary
scalar \EE spectrum by the defect perturbations is likely to be
insignificant relative to that produced by the inflationary
perturbations, since defects are sub-dominant. It is also likely to be
negligible relative to the B-mode from defects since defects
still produce a strong B-mode signal even when their sub-dominance is
taken into account. The lensing of the defect \EE and \BB spectra by the
inflationary perturbations should be negligible also since defects
contribute equally strongly to these two spectra.

With the temperature spectra all normalized to the same value at $\ell=10$, the
polarization spectra are all of similar magnitude and shape (if
normalized to the same $G\mu$, the Abelian Higgs contribution would be
much higher).  In the \EE case, all three defect models peak at the
same value of $\ell\sim1000$, with the semilocal component having
slightly less pronounced oscillations, and the secondary peak at low
$\ell$ coming from reionization follows the following trend: it peaks
at the lowest value of $\ell$ for textures, next for semilocal
strings, and highest for Abelian Higgs strings. The same trend can be
seen for the low-$\ell$ features of the \TE spectra. The high-$\ell$
behaviour of the \TE is roughly the same for all three cases, showing
large oscillations that eventually die out. Note, however, that the
oscillations for the semilocal case are the smallest, and there is no
alternation between anticorrelation and correlation due to the
combined effect of the scalar--vector--tensor contributions which have
different phases.

The \BB power spectrum is arguably the most interesting prediction
from defects \cite{Pen:1997ae,Seljak:2006hi,Bevis:2007qz,Pogosian:2007gi}: inflation
contributes weakly to the \BB polarization \cite{Kamionkowski:1996ks},
and the signal predicted from defects might be detected in the near
future, providing a smoking gun for the existence of defects in
cosmology. In Fig.~\ref{polar} it can be seen that the shape of the
\BB power spectra is similar for semilocal strings, textures and
Abelian Higgs string. Both peaks follow the aforementioned trend, with
textures peaking for lowest $\ell\sim240$, then semilocal strings $\ell\sim360$, and
finally Abelian Higgs strings $\ell\sim590$; furthermore, the \BB power spectrum for
textures decays first, then that for semilocal strings, and the
Abelian Higgs string \BB power spectrum decays at the highest $\ell$.

\begin{table*}[t]
\begin{ruledtabular}
\begin{tabular}{l|cccc}
Parameter & PL                & PL+SL             & PL+AH
& PL+TX \\ \hline
$\fd$     & $0$               & $0.14\pm0.06$     & $0.09\pm0.05$
& $0.18\pm0.08$     \\
$\ns$     & $0.96\pm0.02$   & $1.02\pm0.04$   & $1.00\pm0.03$
& $1.04\pm0.05$   \\
$h$       & $0.74\pm0.03$     & $0.85\pm0.06$     & $0.84\pm0.06$
& $0.87\pm0.08$    \\
$\Obhh$   & $0.0223\pm0.0007$ & $0.0263\pm0.0021$ & $0.0258\pm0.0020$
& $0.0264\pm0.0026$ \\
$\Omhh$   & $0.126\pm0.007$   & $0.122\pm0.007$   & $0.122\pm0.007$
& $0.118\pm0.008$   \\
$\ln(10^{10} A_s^2)$     & $3.09\pm0.06$     & $2.96\pm0.08$     & $2.99\pm0.08$
& $2.94\pm0.09$     \\
$\tau$    & $0.09\pm0.03$     & $0.11\pm0.04$     & $0.11\pm0.04$
& $0.12\pm0.04$
\end{tabular}
\end{ruledtabular}
\caption{Mean and standard deviations of the power-law model
parameters, using \CMB data only. }
\label{t:bestfitCMB}
\end{table*}

\section{Constraints from current \CMB data}
\label{MCMC}

\begin{figure}[t]
\begin{center}
\includegraphics[width=8cm]{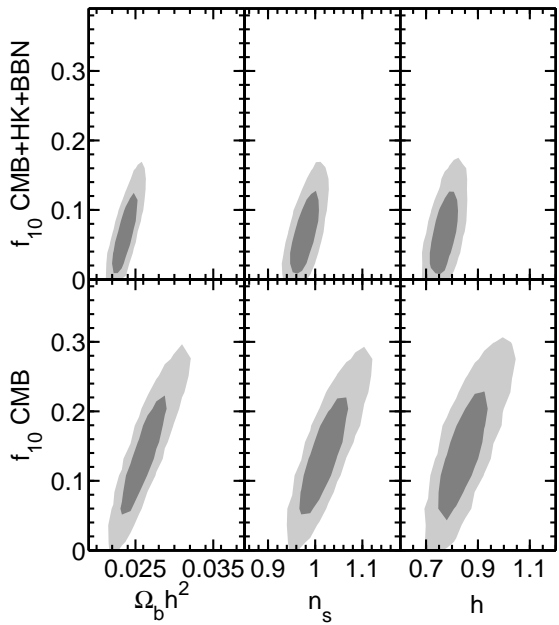}
\caption{\label{figmcmc} The 2-dimensional constraints from
the \mcmc chains using \CMB only data (bottom), and using
\textsc{CMB}+\textsc{BBN}+\HK data (top).}
\end{center} \end{figure}

Clearly, a string component at the $G\mu_{10}$ level shown in
Table~\ref{gmu10} is ruled out by the experiments. Therefore, in order
to quantify the allowed level for a string component, we introduce a
parameter $\fd$ \cite{Bevis:2007gh}: the fractional contribution of
strings to the temperature power spectra at multipole $\ell=10$. We
then use a modified version of CosmoMC \cite{Lewis2002} to perform a multi-parameter Markov Chain Monte Carlo (\textsc{MCMC})
likelihood analysis for \CMB data (\textsc{WMAP}, ACBAR, BOOMERANG, CBI
and VSA projects \cite{\cmbdata}) when semilocal strings or textures are included.
We also reproduce the results for Abelian Higgs strings from \cite{Bevis:2007gh} for comparison. The
likelihood analysis is performed for the usual six-parameter Power-Law
(PL) model ($\Obhh$, $h$, $\ns$, $\tau$, $\Omhh$, $\As$), plus the new
extra $\fd$.  We vary only the normalization of the string (or texture) power
spectra and keep the form of them fixed, following Refs.~\cite{Bevis:2004wk,Bevis:2007gh}. 
This approach is justified since the strings (or textures) contribute only a small fraction to the
total power, and the changes in the cosmological parameters are not large enough that the changes in the form of the string or texture spectra are relevant. The best-fit values of the parameters can be found in
Table~\ref{t:bestfitCMB} for PL alone, and for PL with semilocal (SL)
strings, Abelian Higgs (AH) strings and textures (TX).

Table~\ref{t:bestfitCMB} shows that, as in previous results
\cite{Bevis:2004wk,Bevis:2007gh}, the degeneracies of the cosmological
parameters with respect to the \CMB data allow for a rather high value
of $\fd=0.14\pm0.06$, which corresponds to $G\mu=[1.9\pm0.4]\times
10^{-6}$.  In order to accommodate the defect contribution, other
parameters (most notably $\Obhh$, $h$, and $\ns$) get shifted to
higher values, as was the case for the Abelian Higgs strings
\cite{Bevis:2007gh}.

\begin{figure}[t]
\begin{center}
\includegraphics[width=7cm]{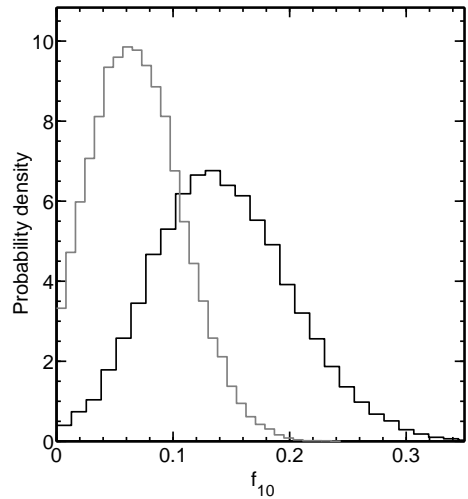}
\caption{\label{1d} The 1-dimensional marginalized likelihood plots for
the fraction of semilocal strings $\fd$ allowed at multipole $\ell=10$
from the \mcmc chains using \CMB only data (dark), and using
\textsc{CMB}+\textsc{BBN}+\HK data (light).}
\end{center} \end{figure}

Figure \ref{figmcmc} shows the marginalized 2D likelihood
distributions for the parameters $\Obhh$, $h$, and $\ns$ versus $\fd$.
It is clear in the figure that there is ample space for a noticable
value of $\fd$. In fact, it can be seen in the 1-D marginalized
likelihood plot (Fig.~\ref{1d}) that the \CMB data alone prefers to have
semilocal strings at a 2-$\sigma$ level.

However, as mentioned before (see Table~\ref{t:bestfitCMB}), the best-fit values
of $\Obhh$ and $h$ are high compared to the concordance model:
Non-\CMB experiments for those two quantities yield $h=0.72\pm0.8$
(the Hubble Key Project \HK \cite{Freedman:2000cf}) and
$\Obhh=0.0214\pm0.0020$ (measurements of deuterium abundance in
high-redshift gas clouds together with big bang nucleosynthesis \BBN
\cite{Kirkman:2003uv}). If we include these two non-\CMB experimental
values into our \mcmc scheme (as gaussian likelihoods), we confirm that
the new set of data is more constraining also for semilocal strings
(as seen in Figs.~\ref{figmcmc} and \ref{1d}). However, the 95\%
confidence level upper bound on $\fd$ for semilocal strings remains
rather high, going from $G\mu<2.6\times10^{-6}$ ($\fd<0.25$) for the
\CMB only case to $G\mu<2.0\times10^{-6}$ ($\fd<0.14$).

We may now use these upper bounds to compare the strengths of the
power spectra of different models. Table~\ref{t:boundCMB} summarises
the values of $\fd$ and $G\mu$ we obtain for our analysis of semilocal
strings and textures, together with that for Abelian Higgs strings
from Ref.~\cite{Bevis:2007gh}.  The parameter 
$\fd$ depends mainly on the form of
the defect power spectra, and we see that the data allow for a higher
contribution from semilocal strings (and textures) than from Abelian Higgs
strings. Besides, the values of $G\mu_{10}$ are also higher for semilocal
strings (and textures) than for the Abelian Higgs strings. Therefore, the
expectation that the constraints on the Abelian Higgs string tensions would be
alleviated by semilocal strings is confirmed: the 95\% upper bound limit for
semilocal strings is $G\mu<2.0\times10^{-6}$ for \textsc{CMB}+\textsc{BBN}+\HK 
($G\mu<2.6\times10^{-6}$ for \CMB only) whereas for Abelian Higgs string it is
$G\mu<0.9\times10^{-6}$ for \textsc{CMB}+\textsc{BBN}+\HK
($G\mu<1.1\times10^{-6}$
for \CMB only).

\begin{figure}[t]
\begin{center}
\includegraphics[width=8cm]{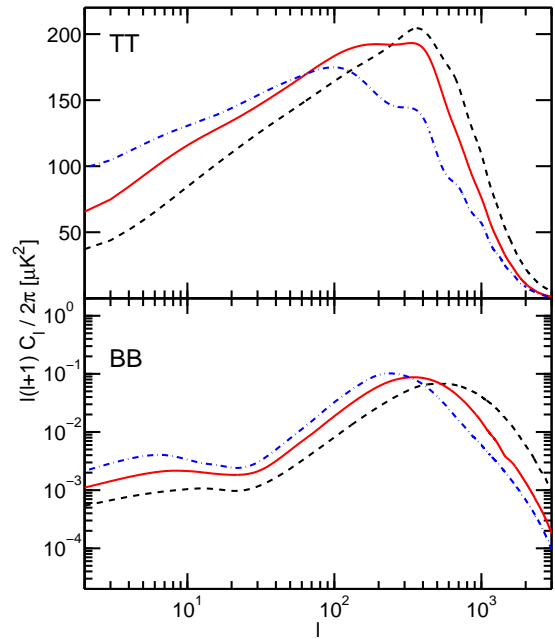}
\vspace{0cm}\caption{\label{pol2} Temperature power spectra (top) and
\BB polarization power spectra (bottom) for semilocal strings (solid
red), compared to Abelian Higgs strings (dashed black) and textures
(dot-dash blue). The curves are normalized using the 95\% confidence
level upper bound from a \mcmc analysis using \CMB, \BBN and \HK
data.}
\end{center} \end{figure}

Figure \ref{pol2} shows the temperature power spectrum and the \BB
polarization spectrum for semilocal strings, Abelian Higgs strings and
textures, normalized by using the upper bounds obtained from
performing the \mcmc likelihood analysis for both \CMB and non-\CMB
(\BBN and \textsc{HKP}) data.  In the temperature power spectrum figure it can
be seen that due to the different shapes of the curves, even though
the textures are allowed the highest level of power at $\ell=10$, the
semilocal and Abelian Higgs strings go up to higher peaks. The signal
is highest for textures for $\ell\lesssim 60$, for semilocal strings
for $60\lesssim\ell\lesssim 250$ and then for Abelian Higgs for
$250\lesssim\ell$.
 On the other hand, for the \BB polarization, the
normalization makes the trend of the power coming from the three cases just 
the reverse of the temperature case: textures have the highest peak, followed
by semilocal strings and then Abelian Higgs strings.
 Needless to say, the normalization
does not change the value of $\ell$ at which each case peaks, and we
find that textures peak first, then semilocal strings, and then
Abelian Higgs strings.

It is worth mentioning that, in all cases studied, the value for the
spectral index $\ns$ is higher than the one obtained for cases without 
defects; and $\ns=1$ is not ruled
out\footnote{See Ref.~\cite{Liddle:2006} for a discussion about $\ns$
in light of WMAP3 from a Bayesian model selection point of view.} (see
table~\ref{t:boundCMB}).  Therefore, when comparing different
cosmological models, one should also consider the (scale-invariant)
Harrison--Zel'dovich (HZ) model with 5 parameters ($\Obhh$, $h$,
$\tau$, $\Omhh$, $\As$) and a fixed $\ns=1$.

\begin{table}[t]
\begin{ruledtabular}
\begin{tabular}{l|ccc|ccc}
      & \multicolumn{3}{c|}{\CMB only} & \multicolumn{3}{c}{
\textsc{CMB}+\textsc{BBN}+\HK} \\
Model & $\fd$ & $G\mu$ &$n_s$~ & $\fd$ & $G\mu$ & $n_s$~  \\\hline
SL & $0.25$ & $2.6\times 10^{-6}$ & $1.09 $~  & $0.14$ & $2.0\times
10^{-6}$ & $1.01$~  \\
TX & $0.33$ & $2.5\times 10^{-6}$ & $1.14$~  & $0.16$  & $1.8\times
10^{-6}$ & $1.02$~  \\
AH & $0.17$ & $1.1\times 10^{-6}$ & $1.06$~  & $0.10$  & $0.9\times
10^{-6}$ & $1.00$~ \\
\end{tabular}
\end{ruledtabular}
\caption{95\% upper bounds on $\fd$, $G\mu$ and $n_s$ for  PL+X,
  using \CMB only, and using \textsc{CMB}+\textsc{BBN}+\textsc{HKP}.} 
\label{t:boundCMB}
\end{table}

\begin{table}[t]
\begin{ruledtabular}
\begin{tabular}{l|rr|rr}
      & \multicolumn{2}{c|}{\CMB only} & \multicolumn{2}{c}{\textsc{CMB}+\textsc{BBN}+\textsc{HKP}} \\
Model & $\Delta\chi^2$ & $\ln(E)$~ & $\Delta\chi^2$ & $\ln(E)$~  \\\hline
HZ    & $6.6$  & $-0.6\pm0.1 $~ & $8.4$  & $-1.4\pm0.1$~ \\
PL    & $0$    & $0$~          & $0$    & $0$~          \\
HZ+SL & $-4.4$ & $2.5\pm0.2$~  & $-0.7$ & $1.0\pm0.3$~  \\
PL+SL & $-4.5$ & $1.0\pm0.3$~  & $-1.9$ & $-1.0\pm0.1$~ \\
HZ+AH & $-3.5$ & $1.8\pm0.2$~  & $1.4$  & $-0.7\pm0.2$~ \\
PL+AH & $-3.5$ & $-0.1\pm0.2$~  & $-1.4$ & $-1.8\pm0.2$~ \\
HZ+TX & $-3.2$ & $2.0\pm0.3$~  & $-0.5$  & $0.3\pm0.3$~  \\
PL+TX & $-3.4$ & $1.2\pm0.3$~  & $-1.2$ & $-1.2\pm0.1$~ \\
\end{tabular}
\end{ruledtabular}
\caption{Goodness of fit and Bayesian Evidence relative to the
  concordance model PL.} 
\label{t:evidenceCMB}
\end{table}

Table~\ref{t:evidenceCMB} shows the goodness of fit for the
concordance PL model, as well as for the HZ model, when Abelian Higgs
strings, semilocal strings and textures are added; for both \CMB only
and \textsc{CMB}+\textsc{BBN}+\HK data. In most cases, the inclusion of the defects
makes the fit better, with semilocal strings obtaining marginally the
best fit. This is not too surprising for the cases where defects are
added to the PL model, because we are adding an extra parameter to the
PL model, and of course obtaining a better fit. But the models with HZ +
defects have the same number of parameters as the PL model (6
parameters), and surprisingly, the fit is better when considering \CMB
data only.

In order to compare different models, we use Bayesian evidence values
\cite{Liddle:2006tc}. The evidences are calculated using the
Savage--Dickey method \cite{Kunz:2006mc,Trotta:2007hy} with flat
priors of $0<\fd<1$ and $0.75<n_s<1.25$ (Table~\ref{t:evidenceCMB}). In
the case of \CMB data only, adding a defect contribution to the PL
model makes little change to the evidence, whereas HZ+defects models
have a significant higher evidence (and semilocal strings are
marginally better than textures or Abelian Higgs strings). Therefore,
taken at face value, HZ+SL strings is the six-parameter model which
fits the data best, though the others are not convincingly worse.  With the
non-\CMB data added, the evidences get reduced, but not to the level
to conclude that a string contribution is ruled out.

This type of comparison will have some dependence on the choice of
prior used for the defect component (the other priors are less
important as they are shared by the models). 
Our imposition of a prior on $\fd$ is just one out of several plausible
options. A prior on $\mu$ might be more appropriate but is hard to
formulate; for instance a logarithmic (Jeffreys) prior on $\mu$ would
put most of the prior weight on models with negligible defect
contribution and hence leave the data unable to distinguish them (the
same situation occurs with the inflationary tensor component -- see
Ref.~\cite{Liddle:2006}). In any event, we can safely say that present
data are not good enough to exclude a defect contribution, and leave
open the tantalising possibility that such a contribution may be a
significant one.

\section{Discussion}
\label{discussion}

In this article we report on the first ever semilocal string
simulations to predict \CMB power spectra. This has been possible
because of recent improvements in simulating field theoretical
defects, as there is no Nambu--Goto type approximation for semilocal
strings. The case of Abelian Higgs model strings is much better known,
so we compared the semilocal string predictions to the Abelian Higgs
string ones, trying to minimize the differences in the approaches. 
Since semilocal strings were expected to share properties with both an 
Abelian Higgs string prediction and a texture prediction,
we also produced a set of simulations for texture-type defects
to compare all three defect contributions. 

Our first conclusion is that the amplitude of the spectra for the
semilocal strings is indeed lower than in Abelian Higgs models (for a
given energy scale); therefore, the normalization at $\ell=10$ to
match the \WMAP value gives a higher value for semilocal strings
$G\mu_{10}=5.3\times 10^{-6}$ than for Abelian Higgs strings
$G\mu_{10}=2.7\times 10^{-6}$.  Moreover, the form of the power
spectra for semilocal strings allows for their fractional contribution
to the temperature power spectra at multipole
$\ell=10$ to be higher than the one for Abelian Higgs strings: the
95\% confidence upper bounds are $\fd<0.25$ ($\fd<0.14$) for semilocal
strings for \CMB only (\textsc{CMB}+\textsc{BBN}+\textsc{HKP}); whereas $\fd<0.17$
($\fd<0.10$) for Abelian Higgs strings.

These two ingredients contribute to allow a higher value of $G\mu$ for
semilocal strings when \CMB data (and also when \HK and \BBN data) are
taking into account: $G\mu<2.6\times10^{-6}$ ($G\mu<2.0\times10^{-6}$)
for semilocal strings for \CMB only (\textsc{CMB}+\textsc{BBN}+\textsc{HKP}); whereas
$G\mu<1.1\times10^{-6}$ ($G\mu<0.9\times10^{-6}$) for Abelian Higgs strings.
Thus, if a promising high-energy physics inflation model faces a
problem of creating Abelian Higgs cosmic strings at a scale that it is
too high for the \textsc{CMB}, a possible resolution could be to extend the model
to predict semilocal strings instead.  This weakens the $G\mu$ 
constraint by a factor of nearly three, which may give the theory
extra breathing space.

As in the case of textures \cite{Bevis:2004wk} and 
local strings \cite{Battye:2006pk,Bevis:2007gh}, the best fit for the cosmological
parameters when taking into account semilocal strings gives a rather
high spectral index $n_{\rm s}$, comparing to the concordance
model. For \CMB only data, the preferred value of $n_{\rm s}$ is
higher than 1 ($n_{\rm s}=1.02\pm0.04$), whereas if \BBN and \HK data
are included $n_{\rm s}=0.98\pm0.02$. This is in contrast to the
Abelian Higgs data, where the \CMB data alone predict $n_{\rm s}=1$;
but the result is very similar when \BBN and \HK data are included
\cite{Bevis:2007gh}.  This result also alleviates pressure on
high--energy physics models (such as D-term and F-term hybrid
inflation) which predict defects and an $n_s$ close to unity.

As $\ns = 1$ is close to our best-fit values, we studied how well a
combination of Harrison--Zel'dovich plus defects would fit the data,
comparing it to the concordance model.  Using goodness of fit and
Bayesian Evidence criteria, we find that models with defects are not
ruled out (even when \BBN and \HK data are included), and should be
considered as competitive models. In fact, taking into account only
\CMB data, there is significant evidence in favour of a
Harrison--Zel'dovich plus semilocal strings model.

We have not included large-scale structure datasets into our analysis
for two reasons.  Firstly, perturbations from defects are non-linear
and non-Gaussian as soon as they are created \cite{Bevis:2004wk}, and
so it is not clear that the commonly-used model \cite{Cole:2005sx} for
non-linear corrections to the galaxy power spectrum applies. Secondly,
recent analyses \cite{Percival:2006gt,Sanchez:2007rc} show
scale-dependent bias even on relatively large scales ($0.05 < k/h\mathrm{Mpc}^{-1} < 0.15
$), and make clear that more work is needed
before the galaxy power spectrum can be used for robust estimates of
cosmological parameters.

The prediction for polarization power spectra from semilocal strings
is similar to that of Abelian Higgs strings. As a matter of fact, the
height of the \BB polarization peak (normalized using the $95\%$
confidence level upper bound on $\fd$) is of the same order of 
magnitude for
Abelian Higgs strings, semilocal strings and textures, though  
the position at which the maximum occurs vary from case to case.
A detection
of a \BB signal on these scales could be an indicator of the existence
of cosmic defects \cite{Bevis:2007qz}.

In the present work we have studied only BPS semilocal strings, but we
will in a future work extend the analysis to lower values of $\beta$,
which would arise in F-term supersymmetric models (or even in P-term
models \cite{Burrage:2007yu,Burrage:2007bv}). The behaviour of semilocal strings does
depend strongly of $\beta$ \cite{Achucarro:2005tu}: the lower the
value of $\beta$ the more the semilocal strings behave as
Abelian Higgs strings. Tracking how the \CMB predictions change with
respect to $\beta$ will give us insight into how the different
components in a string network contribute to the creation of
anisotropies.

There is also a whole new class of strings coming from string
inflation: $(p,q)$-strings
\cite{Copeland:2003bj,Sarangi:2002yt,Dvali:2003zj}. Even though there
is some work about the networks of these defects
\cite{Saffin:2005cs,Hindmarsh:2006qn,Rajantie:2007hp,Urrestilla:07},
basic properties, such as scaling, are still not really
understood. Provided some sound \CMB predictions from these strings
are calculated, it would be interesting to compare the results with
those of Abelian Higgs or semilocal strings.

Ultimately, one would like to be able to discriminate between all the
different defect types, be it Abelian Higgs, semilocal, textures of
$(p,q)$-strings. As we have shown here, the predictions do look
similar, but there are differences, e.g., the aforementioned position of
the \BB peak. The question is whether these differences are strong
enough, or whether future data will be good enough, to pinpoint
whether defects do appear and if so, which type.  In a future
publication we will address this issue.

\begin{acknowledgments}

We thank M.~Nitta for pointing out Ref.~\cite{Eto:2006db} 
and K.~Sousa for discussion.
We acknowledge support from PPARC/STFC (N.B., M.H, A.R.L.), the Swiss
NSF (M.K.), Marie Curie Intra-European Fellowship MEIF-CT-2005-009628
(J.U.).  This work was partially supported
by Basque Government (IT-357-07), the Spanish Consolider-Ingenio 2010
Programme CPAN (CSD2007-00042) and FPA2005-04823 (J.U.).
The simulations were performed on
\textsc{cosmos}, the UK National Cosmology Supercomputer, supported by
SGI, Intel, HEFCE and PPARC.

\end{acknowledgments}

\appendix

\section{Texture simulations}

In  this work we have compared the \CMB predictions to those from the
Abelian Higgs model and texture models. For the latter we have
performed a new set of simulations for the linear-$\sigma$ global 
texture model comparing the results additionally with previous ones from
Refs.~\cite{Pen:1997ae, Seljak:1997ii, Durrer:1998rw,
Bevis:2004wk}.

We exploited the fact that a  texture model can be obtained directly
from Lagrangian Eq.~(\ref{lagr}) by setting the gauge field to zero,
obtaining 
\bea 
\mathcal{L}
&=&\left|\d_\mu\phi_1\right|^2+\left|\d_{\mu}\phi_2\right|^2-
\frac{\lambda}{4}\left(\left|\phi_1\right|^2 +
\left|\phi_2\right|^2-\eta^2\right)^2\nonumber\\   
&=& \sum_{i=1}^4 (\d_\mu \psi_i)^2 - \frac{\lambda}{4}
\left(\sum_{i=1}^4 \psi_i^2 - \eta^2\right)^2
\label{lagrtex}
\eea
where $\psi_i, \; (i=1...4)$ are real scalar fields and
$\phi_1=\psi_1+i\psi_2$, $\phi_2=\psi_3+i\psi_4$.

\begin{figure}[h]
\begin{center}
\includegraphics[width=8cm]{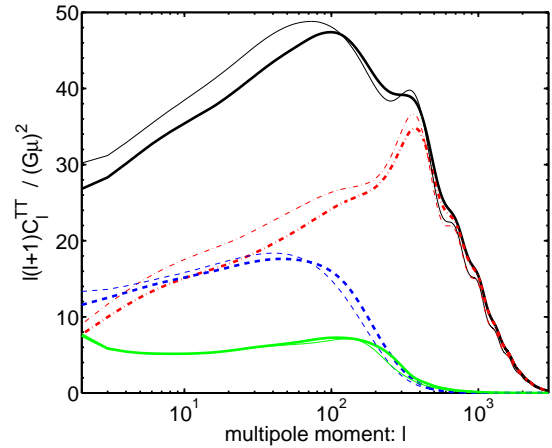}
\caption{\label{gon}The temperature power spectra from linear $\sigma$
textures (thick) obtained and used in this article; together with the
spectra from a non-linear $\sigma$ model (thin) from \cite{\longprd}. The different
components are scalar (dash-dot red), vector (dashed blue) and tensor
(lower solid green).}
\end{center}
\end{figure}

This is the Lagrangian we have simulated to obtain the figures used in
the main body if this article. Note that it is not canonically
normalized with respect to the real scalar fields, which becomes clear
when compared to the non-linear $\sigma$ model: 
\be 
\mathcal{L}=\half
\sum_{i=1}^4 (\d_\mu \tilde\psi_i)^2 - \tilde\lambda \left(\sum_{i=1}^4
\tilde\psi_i^2 - \kappa^2\right)
\label{cano}
\ee where in this case $\tilde\lambda$ is a Lagrange multiplier fixing
the fields onto the vacuum manifold.

The normalization of the textures is usually given for the Lagrangian
canonically normalized for scalar fields as \cite{Durrer:2001cg}
\be 
\varepsilon=4\pi G\kappa^2\,.  
\ee 
Comparing both Lagrangians (\ref{lagrtex}) and (\ref{cano}), it can be
seen that $\kappa^2=2\eta^2$, so the normalization scheme for textures
used throughout this work Eq.~(\ref{mu}) relates to that of
Ref.~\cite{Durrer:2001cg} by \be \varepsilon=4G\mu \ee Note that
Ref.~\cite{Pen:1997ae} uses another definition of $\varepsilon$,
bigger by a factor $2\pi$.

The procedure followed to obtain the \CMB predictions from textures
was exactly the same as for the semilocal case (and for the Abelian Higgs
case). The results can be seen in Fig.~\ref{SVT}. Earlier work
on global textures \cite{Crittenden:1995xf,Durrer:1995sf,Durrer:1998zj} represented important steps towards the formalism
we are using here. However, they neglected decoherence and so overemphasized 
the presence of the acoustic peaks.

We also compared the texture predictions we obtain in our simulations
with those of a non-linear $\sigma$ model by some of the authors of
this article \cite{\longprd} in Fig.~\ref{gon}. Those simulations were performed
using different parameters; e.g., smaller lattices $(256^3)$ and smaller
number of eigenvectors used to obtain the \CMB power spectra. In any
case, the agreement with our linear-$\sigma$ model is very good, which
makes our comparison of texture with semilocal and Abelian Higgs strings more
widely valid.

\bibliography{CMBsemilocal}

\end{document}